\documentclass[11pt, a4paper]{article}

\usepackage{setspace}
\usepackage{mathtools,amsmath,amssymb,amsthm,color}
\usepackage{mathrsfs,bm,comment,bbm}
\usepackage{float}
\usepackage{cases}
\usepackage{algorithm, algorithmicx, algpseudocode}
\usepackage{comment}

\usepackage[pdftex]{graphicx}
\usepackage {booktabs}
\usepackage{multirow}

\usepackage{natbib}

\theoremstyle{plain}

\theoremstyle{definition}
\newtheorem{rem}{Remark}[section]

\definecolor{darkblue}{rgb}{0.0,0.0,0.55}
\RequirePackage[colorlinks,citecolor=darkblue,urlcolor=darkblue,linkcolor=darkblue]{hyperref} 
\usepackage{xr} 

\makeatletter
 
  \@addtoreset{equation}{section}
\makeatother

\newcommand{\argmin}{\mathop{\rm argmin}\limits}

\usepackage[top=30mm,bottom=27mm,left=30mm,right=30mm]{geometry}

  \allowdisplaybreaks
 
\begin{document}

\title{Bayesian Boundary Trend Filtering}
\author{Takahiro Onizuka, Fumiya Iwashige and Shintaro Hashimoto\\
Department of Mathematics, Hiroshima University, Japan
}

\maketitle

\begin{abstract}
Estimating boundary curves has many applications such as economics, climate science, and medicine. Bayesian trend filtering has been developed as one of locally adaptive smoothing methods to estimate the non-stationary trend of data. This paper develops a Bayesian trend filtering for estimating the boundary trend. To this end, the truncated multivariate normal working likelihood and global-local shrinkage priors based on the scale mixtures of normal distribution are introduced. In particular, well-known horseshoe prior for difference leads to locally adaptive shrinkage estimation for boundary trend. However, the full conditional distributions of the Gibbs sampler involve high-dimensional truncated multivariate normal distribution. To overcome the difficulty of sampling, an approximation of truncated multivariate normal distribution is employed. Using the approximation, the proposed models lead to an efficient Gibbs sampling algorithm via the P\'olya-Gamma data augmentation. The proposed method is also extended by considering a nearly isotonic constraint. The performance of the proposed method is illustrated through some numerical experiments and real data examples. 

\end{abstract}

\noindent
{\bf Keywords}: Boundary trend; Gibbs sampler; horseshoe prior; nearly isotonic regression; trend filtering

\section{Introduction}
\label{sec:1}

Consider the nonparametric regression model
\begin{align}\label{regression-model}
y_i=f(x_i) +\varepsilon_i,\quad i=1,\dots,n,
\end{align}
where $y_1,\dots,y_n \in \mathbb{R}$ are outcomes, $x_1,\dots,x_n \in \mathbb{R}$ are input points, $f$ is the underlying function to be estimated, and $\varepsilon_1,\dots,\varepsilon_n$ are independent errors which are not assumed to be centered, but to have one-sided support such as $(-\infty,0]$. The regression function $f$ describes the support frontier. For such models, some of the regularity conditions on the statistical model may be violated. For example, \cite{S94} considered the linear regression model with a class of one-sided error distributions including the exponential and Weibull distributions. In this case, the support of the distribution of the response variable depends on unknown parameters. He also showed that the asymptotic distribution of the estimator of the regression coefficient vector is non-normal. For this reason, the model \eqref{regression-model} is generally called a ``non-regular model". In the context of nonparametric regression, some authors proposed the methods and showed their theoretical properties for the model defined by \eqref{regression-model} \citep[see e.g.][]{DST84, HPS98, HS02, DNP16, RS17, RS20b, RS20a, STM22}. 
In particular, \cite{DLN17} provided an excellent {\tt R} package to implement some nonparametric boundary estimation methods including shape restriction and robust estimation. There are many applications of the model \eqref{regression-model}, for instance, in the microeconomic theory where the support boundary is considered as the set of the most efficient businesses or industries, in climate science where the trend of maximum values of temperature is important to come up with an environmental policy. Furthermore, these considerations often lead to the assumption of monotonicity/nearly monotonicity. In nonparametric regression or nonparametric curve fitting problems, shape constraints such as monotonicity and convexity are also useful when we have some prior information on the shape of data \citep[see e.g.][]{Robertson88}. For boundary curve estimation, \cite{DNP16} proposed spline smoothing methods under monotone/concave constraint, and provided an efficient optimization algorithm. 
We note that boundary regression models relate to nonparametric quantile regression with high or low quantile levels. The method is also called extremal quantile regression, and it might be interpreted as an exploratory tool rather than as a method for final boundary analysis. In this paper, we are interested in the boundary regression curve, not the extremal quantile regression curve.

To estimate unknown regression function, trend filtering is one of the popular methods which was originally proposed for estimating a nonparametric regression problem \citep{KKBG09}. \cite{T14} showed some theoretical properties of trend filtering to estimate the mean trend of data, and \cite{RT16} provided a fast and efficient optimization algorithm to calculate the estimates. Some variations were also provided by \cite{RT16} such as shape constraint, robustness against outliers, and so on. More recently, trend filtering has been widely extended in various directions such as quantile regression \citep{BGC20} and functional data analysis \citep{Wakayama21}.  
On the other hand, smoothing methods via trend filtering have also been developed in Bayesian analysis \citep{R15, FM18, OHS22, Wakayama22, Heng22}. While the trend filtering method was originally a nonparametric regression problem, we often assume the parametric (working) likelihood function and prior distribution. The Bayesian approach has some advantages: 1) capable of full probabilistic uncertainty quantification through posterior distribution, 2) flexible shrinkage by using global-local shrinkage priors \citep{CPS10}, and 3) estimating regularization parameter from Markov chain Monte Carlo method. 

In this paper, we develop a locally adaptive boundary smoothing method using Bayesian trend filtering. To this end, we introduce the truncated multivariate (tMVN) distribution as a working likelihood and shrinkage priors for differences. Using the scale mixtures of normal priors, we can easily derive Gibbs sampling algorithms, while one of the full conditional distributions is the tMVN distribution. It is well-known that sampling from the tMVN distribution is quite challenging even if the dimension of a parameter is moderately large (e.g. one hundred). To overcome this difficulty, we employ an approximation of the indicator function in the tMVN distribution by using the sigmoid function with a tuning scale parameter. The idea comes from the paper \cite{SBP18} on the Bayesian shape constraint regression \citep[see also][]{RPB20}. While they used the approximation for prior distribution, we adopt one to the likelihood function. Using such approximation, we provide an efficient Gibbs sampling algorithm using the P\'olya-Gamma data augmentation \citep{PSW13}. As shrinkage priors, the horseshoe \citep{CPS10}, Laplace, and normal priors are adopted as in \cite{FM18} and \cite{OHS22}. The Laplace prior is also called the Bayesian lasso prior which corresponds to $L_1$-penalty in the original trend filtering, while the horseshoe prior is known as a more flexible shrinkage prior in sparse Bayesian estimation by introducing global and local shrinkage parameters. We also extend the proposed model to the case of some shape constraints. In particular, we propose a nearly isotonic regression method for estimating support boundaries. The nearly isotonic regression which was proposed by \cite{THT11} is a kind of generalization of the original isotonic regression because the method allows violations of monotonicity at some change points. In other words, the nearly isotonic regression is a penalized version of isotonic regression, and it has robustness against structural misspecification for the assumption of monotonicity. We illustrate the performance of the proposed method through some numerical experiments including real data examples and also provide the sensitivity analysis for selecting a tuning parameter of the sigmoid function. 

The rest of the paper is organized as follows. In Section \ref{sec:2}, we formulate the proposed method and prior specification. An efficient Gibbs sampling algorithm using the approximated likelihood and data augmentation is also introduced. Some simulation studies and real data applications are given in Sections \ref{sec:3} and \ref{sec:4}, respectively. R code implementing the proposed methods is available at GitHub repository: 
\begin{center}
\url{https://github.com/Takahiro-Onizuka/BBTF}
\end{center}

\section{Bayesian boundary trend filtering}
\label{sec:2}

\subsection{Review of trend filtering}
The original $\ell_1$ trend filtering was proposed by \cite{KKBG09} to estimate nonparametric regression model $y_i=f(x_i) +\varepsilon_i$ for $i=1,\dots,n$. The trend filtering estimate is given by solving the following optimization problem: 
\begin{align}\label{TF}
\min_{\theta \in \mathbb{R}^n} \sum_{i=1}^n (\theta_i-y_i)^2 +\lambda\|D_n^{(k+1)}\theta\|_1,
\end{align}
where $y=(y_1,\dots,y_n)^{\top}$, $\theta=(\theta_1,\dots,\theta_n)^{\top}=(f(x_1),\dots,f(x_n))^{\top}$, $\lambda>0$ is a tuning parameter which controls the smoothness of the trend, and $D^{(k+1)}$ is a $(n-k-1)\times n$ difference operator matrix of order $k+1$ defined by
\begin{align*}
    D_n^{(1)}=\left(
    \begin{matrix}
     1 & -1 & &\\
     & \ddots& \ddots & \\
     & & 1&-1 
    \end{matrix}
    \right)\in \mathbb{R}^{(n-1)\times n},\quad 
    D_n^{(k+1)}=D_{n-k}^{(1)}D_n^{(k)}.
\end{align*}
The order $k$ is the order of the piecewise polynomials. For example, $k=0,1$ and $2$ correspond to piecewise constant, piecewise linear, and piecewise quadratic, respectively. The trend is also called the fused lasso for $k=0$ in \eqref{TF}. 
This optimization problem provides the estimate of the mean trend, while an extension to the quantile one is provided by \cite{BGC20} using the check loss function instead of the squared loss function. Theoretical properties of quantile trend filtering have been shown by \cite{PS20}. We note that the purpose of trend filtering is to estimate “the value of the underlying function at the data points” and not to estimate the function itself. In other words, the estimate from trend filtering is not a function and it has a different purpose than the spline methods from which the function is obtained as an estimate.

On the other hand, Bayesian trend filtering methods have been developed in recent years. A Bayesian alternative to trend filtering is formulated by the sequence model
\begin{align}\label{Bayesian-TF}
y_i=\theta_i + \varepsilon_i, \ \ \varepsilon_i\sim p(\cdot), \ \ D_n^{(k+1)}\theta \sim \pi(\cdot), \quad i=1,\dots, n,
\end{align}
where $\varepsilon_1,\dots,\varepsilon_n$ are independent errors usually assumed the normal distribution as a working likelihood, and we assume a class of shrinkage priors for the difference $D_n^{(k+1)}\theta$ \citep[see. e.g.][]{R15, FM18, Heng22}. Note that if we assume the Laplace prior or Bayesian lasso prior for $D_n^{(k+1)}\theta$, the resulting maximum a posteriori (MAP) estimate is the same as that of the solution in the problem \eqref{TF}. Bayesian quantile trend filtering was proposed by \cite{OHS22fast} using the asymmetric Laplace density for $p(\cdot)$ in \eqref{Bayesian-TF}. \cite{OHS22fast} also provided a calibration algorithm using the variational Bayes method to obtain valid credible intervals under possibly misspecified asymmetric Laplace likelihood. Although existing methods provide a reasonable estimate for mean or quantile trend, we can not apply such methods to nonparametric regression models with one-sided error as \eqref{regression-model}.

To this end, we consider the following optimization problem to estimate the boundary trend: 
\begin{align}\label{BTF}
\min_{\theta \ge y} \sum_{i=1}^n (\theta_i-y_i)^2 +\lambda\|D_n^{(k+1)}\theta\|_1,
\end{align}
where the relation $\theta \ge y$ between two vectors $y=(y_1,\dots,y_n)^{\top}$ and $\theta=(\theta_1,\dots,\theta_n)^{\top}$ implies $\theta_i \ge y_i$ for all $i$. Note that the constraint $\theta \ge y$ leads to the estimation of the upper boundary trend. 
When we estimate the lower boundary trend, we may consider the constraint $\theta \le y$ instead of $\theta \ge y$. To the best of our knowledge, such boundary trend filtering has not been proposed in terms of trend filtering. In particular, we focus on Bayesian boundary trend filtering. As mentioned in Section \ref{sec:1}, the Bayesian approach has several attractive properties for sparse estimation including trend filtering. For these reasons, we focus on the Bayesian boundary trend filtering in this paper. We also see the advantages such as locally adaptive smoothing and uncertainty quantification through real data analysis in Section \ref{subsec:4.2}.

\subsection{Bayesian boundary trend filtering} \label{subsec:2.2}

Following the model \eqref{Bayesian-TF} and \eqref{BTF}, we formulate the Bayesian boundary trend filtering. Without loss of generality, we only consider the estimation of the upper boundary trend. 
For each $y_i$ ($i=1,\dots,n$), we consider the following model:
\begin{align}\label{model}
y_i=\theta_i + \varepsilon_i,\quad \varepsilon_i \sim \mathrm{HN}(0, \sigma^2)
\end{align}
where $\mathrm{HN}(\mu, \sigma^2)$ represents the (upper truncated) half-normal distribution with location parameter $\mu$ and scale parameter $\sigma^2$. From \eqref{model}, the probability density function of $y_i$ given $\theta_i$ is
\begin{align*}
p(y_i\mid \theta_i,\sigma^2)=\sqrt{\frac{2}{\pi \sigma^2}}\exp\left(-\frac{1}{2\sigma^2}(y_i-\theta_i)^2\right) 1_{\{y_i\le \theta_i\}}(y_i), \quad i=1,\dots,n,
\end{align*}
where $1_{A}(x)$ is an indicator function defined by $1_{A}(x)=1$ if $x \in A$ and $1_{A}(x)=0$ otherwise. The corresponding likelihood function of $\theta$ is given by
\begin{align}\label{likelihood}
L(\theta,\sigma^2 \mid y)\propto (\sigma^2)^{-n/2}\exp\left(-\frac{1}{2\sigma^2}\sum_{i=1}^n (y_i-\theta_i)^2\right)\prod_{i=1}^n 1_{\{y_i\le \theta_i\}}(\theta_i),
\end{align}
which is that of the truncated multivariate normal distribution restricted to the region $\mathcal{C}=\{\theta \in \mathbb{R}^n \mid y_i\le \theta_i,\ i=1,\dots, n\}$. 

Next, we introduce shrinkage priors on differences. We define the $k+1$ th order difference operator $D$ as
\begin{align*}
    D=\left(
    \begin{matrix}
     I_{k+1} & O\\
     \multicolumn{2}{c}{D_n^{(k+1)}}
    \end{matrix}
    \right),
\end{align*}
where $I_{k+1}$ is $(k+1)\times (k+1)$ identity matrix, and $O$ is zero matrix. 
We consider shrinkage priors on $D\theta$. The shrinkage priors we consider here are the horseshoe, Laplace, and normal priors \citep[see also][]{FM18, OHS22fast}. The horseshoe and Laplace priors can be represented by the scale mixtures of normal distribution: 
\begin{align}\label{prior--Dtheta}
D\theta\mid\tau^2,\sigma^2,u \sim N_n(0,\sigma^2 U),
\end{align}
where $U=\mathrm{diag}(u_1^2,\dots,u_{k+1}^2,\tau^2u_{k+2}^2, \dots, \tau^2u_{n}^2)$ with shrinkage parameters $u_i$ and $\tau$. Here, $\tau$ and $u_i$  are called global and local parameters respectively, and this formulation enables locally adaptive smoothing. In \eqref{prior--Dtheta}, we employ a prior distribution on $D\theta$ which depends on error variance $\sigma^2$. The advantage of such a conditional prior is that the scale of the prior is automatically adjusted when we change units of observations. Such a formulation of the prior is widely used when we assume the normal prior \citep[e.g.][]{PS12}. Since the matrix $D$ is non-singular, the prior of $\theta$ can be rewritten as
\begin{align}\label{prior--theta}
    \theta\mid\tau^2,\sigma^2,u \sim N_n(0,\sigma^2(D^{\top}U^{-1} D)^{-1}).
\end{align}
We note that prior distributions for $\tau$ and $u_i$ depend on the shrinkage priors. Shrinkage priors we consider in the paper are expressed as the marginal priors of the following hierarchical priors.
\begin{itemize}
\item Horseshoe prior:
\begin{align*}
& \theta\mid\tau^2,\sigma^2,u \sim N_n(0,\sigma^2(D^{\top}U^{-1} D)^{-1}),\ \mathrm{where}\ U=\mathrm{diag}(u_1^2,\dots,u_{k+1}^2,\tau^2u_{k+2}^2, \dots, \tau^2u_{n}^2)\\  
&\tau \sim C^+(0,1), \ \ u_i^2\sim \mathrm{IG}(a_{u_i}, b_{u_i})\ (i=1,\dots,k+1), \ \ u_i \sim C^+(0,1)\ (i=k+2,\dots,n)
\end{align*}
\item Laplace prior:
\begin{align*}
& \theta\mid\sigma^2,u \sim N_n(0,\sigma^2(D^{\top}U^{-1} D)^{-1}),\ \mathrm{where}\ U=\mathrm{diag}(u_1^2,\dots,u_{k+1}^2,u_{k+2}^2, \dots, u_{n}^2)\\  
&u_i^2\sim \mathrm{IG}(a_{u_i}, b_{u_i})\ (i=1,\dots,k+1), \ \ u_i \sim \mathrm{Exp}(\gamma^2/2)\ (i=k+2,\dots,n),\ \ 
\gamma^2 \sim \mathrm{IG}(a_{\gamma}, b_{\gamma})\end{align*}
\item Normal prior:
\begin{align*}
& \theta\mid\sigma^2,\tau^2,u_1,\dots,u_{k+1} \sim N_n(0,\sigma^2(D^{\top}U^{-1} D)^{-1}),\ \mathrm{where}\ U=\mathrm{diag}(u_1^2,\dots,u_{k+1}^2,\tau^2, \dots, \tau^2)\\  
&u_i^2\sim \mathrm{IG}(a_{u_i}, b_{u_i})\ (i=1,\dots,k+1), \ \ \tau^2 \sim \mathrm{IG}(a_{\tau}, b_{\tau})
\end{align*}
\end{itemize}
Here, $\mathrm{IG}(a,b)$ and $C^+(a,b)$ are the inverse-gamma distribution with shape $a$ and rate $b$, and the half-Cauchy distribution with location $a$ and scale $b$, respectively. 
We also assume the conjugate proper prior for $\sigma^2$ such as $\sigma^2 \sim \mathrm{IG}(a_{\sigma}, b_{\sigma})$ for some hyper-parameters $a_{\sigma}>0$ and $ b_{\sigma}>0$. 

Such formulations of likelihood and prior distributions lead to tractable full conditional distribution for $\theta$ so that we can easily construct an efficient Gibbs sampler. From \eqref{likelihood} and \eqref{prior--theta}, the resulting full conditional distribution of $\theta$ is the truncated multivariate normal (tMVN) distribution. Efficient sampling algorithms for the tMVN distribution have been developed in recent years. For example, \cite{pakman2014} proposed an exact Hamiltonian Markov chain algorithm, and \cite{B17} proposed accept-reject algorithms that create an exact sample from the tMVN distribution. However, 
it is known that sampling from the high-dimensional tMVN distribution is quite challenging even if the dimension of the parameter is moderately large (e.g. $n=100$). To overcome such a sampling problem, we introduce the following approximation of the indicator function in the likelihood function \eqref{likelihood}. The idea comes from the paper by \cite{SBP18}. 
They proposed an approximation of the tMVN distribution via logistic sigmoid function $\sigma_{\eta}(\xi_i)$:
\begin{align}\label{sigmoid}
\mathbbm{1}(\xi_i\ge 0) \approx \sigma_{\eta}(\xi_i)=\frac{e^{\eta \xi_i}}{1+e^{\eta\xi_i}}, \quad i=1,\dots,n.
\end{align}
Then, the approximate truncated multivariate normal likelihood $L_{\eta}(\theta\mid y)$ is represented by 
\begin{align} \label{L-eta}
L_{\eta}(\theta\mid y) &\propto e^{-\frac{1}{2\sigma^2}(y-\theta)^{\top}(y-\theta)}\prod_{i=1}^n \left(\frac{e^{\eta (\theta_i-y_i)}}{1+e^{\eta(\theta_i-y_i)}}\right).
\end{align}
The approximate distribution \eqref{L-eta} is also called the {\it soft} truncated multivariate normal distribution by \cite{SBP18}. We can easily show that the approximate truncated multivariate normal likelihood $L_{\eta}(\theta\mid y)$ converges to the truncated multivariate normal likelihood $L(\theta\mid y)$ as $\eta\to \infty$ in the sense of $L_1$ convergence \citep[see also][]{SBP18}. Hence, the constant $\eta$ controls the accuracy of the approximation, and we recommend a large value for $\eta$.

\begin{rem}[Adjusted difference matrix for irregular grid] \label{rem2.1}
We give some discussion about an extension to the proposed method to the situation where data is observed at irregular grid. It is equal to that the locations of data $x=(x_1,\dots,x_n)$ have the ordering $x_1<x_2<\dots<x_n$ and $x_{j+1}-x_j$ is not constant. When the locations $x\in\mathbb{R}^n$ are irregular and strictly increasing, \cite{T14} proposed an adjusted difference operator for $k\geq 1$
\begin{align*}
D_n^{(x,k+1)}=D_{n-k}^{(x,1)}\mathrm{diag}\left(\frac{k}{x_{k+1}-x_{1}},\dots,\frac{k}{x_n-x_{n-k}}\right)D_{n}^{(x,k)}
\end{align*}
where $D_{n}^{(x,1)}=D_n^{(1)}$ and when $x_1=1,x_2=2,\dots,x_n=n$, $D_n^{(x,k+1)}$ is equal to $D_n^{(k+1)}$ \citep[see also][]{OHS22fast}. The adjusted difference matrix $D$ is also given by
\begin{align*}
    D=\left(
    \begin{matrix}
     I_{k+1} & O\\
     \multicolumn{2}{c}{ D_n^{(x,k+1)}}
    \end{matrix}
    \right).
\end{align*}
\end{rem}

\subsection{Shape constraints}

There are many phenomena for which monotonic or concave constraints are appropriate such as the does-response curve in medicine and the demand curve in economics. The most popular shape-constraint regression method is isotonic regression. Although the isotonic regression is useful in application, the monotone assumption may be violated at a few points in practice. For example, the global warming in climate change \citep{THT11} and geological observations in seismology \citep{M20} indicate violations of the monotonicity. As a penalized isotonic regression, \cite{THT11} proposed nearly isotonic (NI) regression and the corresponding estimate is defined by
\begin{align*}
\hat{\theta}=\argmin_{\theta \in \mathbb{R}^n} \sum_{i=1}^n (y_i-\theta_i)^2 +  \lambda \sum_{i=1}^{n-1}(\theta_i-\theta_{i+1})_+ ,
\end{align*}
where $(x)_+=\max(x,0)$ and $\lambda>0$ is a tuning parameter that controls the violation of the monotone constraint. \cite{RT16} also applied nearly isotonic constraint to trend filtering. 
We now consider the Bayesian boundary trend filtering under the nearly isotonic constraint. First, the optimization problem is formulated as
\begin{align}\label{NI}
\min_{\theta\ge y} \sum_{i=1}^n (y_i-\theta_i)^2  + \lambda_1 \|D_n^{(k+1)}\theta\|_1+ \lambda_2 \sum_{i=1}^{n-1}(\theta_i-\theta_{i+1})_+,
\end{align}
where $\lambda_1, \lambda_2>0$ are tuning parameters. If the constraint $\theta\ge y$ is removed, the optimization problem is nothing but the original nearly isotonic trend filtering proposed by \cite{RT16}. The third term in \eqref{NI} plays a role of nearly isotonic constraint. In other words, we impose a penalty when the monotonicity is violated. In the Bayesian context, the solution of the model \eqref{NI} is equivalent to the posterior mode under the prior which is proportional to 
\begin{align}\label{NI-prior}
\exp\left\{-\lambda_1 \|D^{(k+1)}\theta\|_1\right\}\exp\left\{-\lambda_2 \sum_{i=1}^{n-1}(\theta_i-\theta_{i+1})_+\right\}.
\end{align}
We can easily deal with the first term in \eqref{NI-prior} like \eqref{prior--theta}. 
For the second term in \eqref{NI-prior}, it is useful to employ the variance-mean mixtures representation \citep[see e.g.][]{PS13}
\begin{align*}
a^{-1}\exp\{-2c^{-1} (ax)_+\}=\int_0^{\infty} \phi_1(x\mid -a v, cv) dv,
\end{align*}
where $\phi_n(\cdot \mid a, b)$ represents $n$-dimensional Gaussian density with mean vector $a$ and covariance matrix $b$. Using this identity, we introduce the prior distribution for the nearly isotonic constraint as
\begin{align}
\exp\left\{- \frac{1}{\rho^2\sigma^2}\sum_{i=1}^{n-1}(\theta_i-\theta_{i+1})_+\right\}
&= \prod_{i=1}^{n-1}\int_0^{\infty} \phi_1(\theta_i-\theta_{i+1}\mid - v_i, 2\rho^2\sigma^2v_i) dv_i , \label{prior--constraint}
\end{align}
and then the conditional prior of $\theta$ for nearly isotonic constraint is rewritten as
\begin{align}
p(\theta\mid \rho^2,\sigma^2,v)&=\prod_{i=1}^{n-1}\phi_1(\theta_i-\theta_{i+1}\mid - v_i, 2\rho^2\sigma^2v_i)=\phi_{n-1}(P\theta\mid -v, V), \label{prior--NIconstraint}
\end{align}
where $n-1$ is the length of the vector $P\theta$, $P=D_n^{(1)}$ and $V=\mathrm{diag}(2\rho^2 \sigma^2v_1, \dots, 2\rho^2 \sigma^2v_{n-1})$ with a scale parameter $\rho^2$. Note that the scale parameter $\rho^2$ plays a role of tuning parameter $\lambda_2$ in \eqref{NI}, and we estimate $\rho^2$ from data assuming the prior distribution on $\rho^2$. In our numerical studies, we use a prior $\rho^2\sim \mathrm{IG}(a_{\rho},b_{\rho})$. Introducing latent variables $v_1,\dots,v_{n-1}$, the prior is written by conditional Gaussian distribution, and then the conditional prior of the shape-restricted trend filtering like \eqref{NI-prior} also becomes Gaussian distribution. While we only consider the monotonically increasing condition, if we assume a nearly decreasing or convex, then we may use $P=-D_n^{(1)}\in \mathbb{R}^{(n-1)\times n}$ or $P=D_n^{(2)}\in \mathbb{R}^{(n-2)\times n}$ respectively.

\begin{rem}[Posterior propriety]\label{rem2.2}
The prior defined by the left-hand side of \eqref{prior--constraint} is improper. If we assume proper priors on the remaining parameters, then we can show that the joint posterior distribution is proper because the prior $p(\theta\mid \rho^2,\sigma^2) = \exp\{-1/(\rho^2\sigma^2)\sum_{i=1}^{n-1}(\theta_i-\theta_{i+1})_+\} $ is bounded by 1 for any $\theta$. Hence, the integral
\begin{align} \label{int--theta}
\int p(y\mid \theta, \sigma^2)p(\theta\mid \sigma^2,\tau^2,u)p(\theta\mid \rho^2,\sigma^2) d\theta.
\end{align}
is bounded by the integral of the product of proper density functions.
\end{rem}

\subsection{Markov chain Monte Carlo algorithm}
\label{subsec:2.4}

In this subsection, we construct an efficient posterior computation algorithm via the Markov chain Monte Carlo method. In the proposed model, we can construct a Gibbs sampler. First of all, we consider the sampling of $\theta$ from the posterior distribution. From \eqref{prior--theta}, \eqref{L-eta} and \eqref{prior--NIconstraint}, the full conditional distribution of $\theta$ is given by the following form:
\begin{align*}
\phi_n(\theta\mid \mu_{\theta}, \Sigma_{\theta})\prod_{i=1}^n \left(\frac{e^{\eta (\theta_i-y_i)}}{1+e^{\eta(\theta_i-y_i)}}\right),
\end{align*}
where $\mu_{\theta}$ and $\Sigma_{\theta}$ are some mean vector and covariance matrix. To simplify the sampling, we put $\xi_i=\theta_i-y_i$ and $\xi=(\xi_1,\dots,\xi_n)^{\top}$. Then the full conditional distribution of $\xi$ is given by 
\begin{align}\label{full-xi}
\phi_n(\xi\mid A^{-1}b, A^{-1})\prod_{i=1}^n \left(\frac{e^{\eta \xi_i}}{1+e^{\eta\xi_i}}\right),
\end{align}
where the matrix $A$ and vector $b$ depend on the type of shrinkage priors. 
For sampling of $\xi$, the P\'olya-Gamma data augmentation proposed by \cite{PSW13} can be applied, and then we can sample $\theta$ from the following three steps: 
\begin{itemize}
\item[(i)] Sample latent variables $\omega_i \mid \xi_i \sim \mathrm{PG}(1,\eta\xi_i)$ for $i=1,\dots, n$.
\item[(ii)] Sample $\xi\mid \omega \sim N_n (\mu_{\omega}, \Sigma_{\omega})$, with
\[\Sigma_{\omega}=\left(\eta^2 \Omega + A\right)^{-1},\quad \mu_{\omega}=\Sigma_{\omega}\left(\eta \kappa +b\right),\]
where $\omega_1,\dots,\omega_n$ are latent variables, $\kappa=(1/2,\dots,1/2)^{\top}$, and $\Omega=\mathrm{diag}(\omega_1,\dots,\omega_n)$.
\item[(iii)] Set $\theta=\xi+y$.
\end{itemize}
In step 1, $\mathrm{PG}(b,c)$ is the P\'olya-Gamma distribution with parameter $b>0$ and $c \in \mathbb{R}$ (see Definition 1 in \cite{PSW13}), and sampling from the distribution can be implemented by using {\tt R} package {\tt pgdraw}, for example.

Although we consider three shrinkage priors (horseshoe, Laplace, and normal priors), we only show the full conditional distributions for the horseshoe type prior. Since Gibbs sampling algorithms under the Laplace and normal type priors can also be derived in the same manner, we here omit them.  

Under the horseshoe prior, using the mixtures of inverse-gamma representation of the half-Cauchy distribution, all full conditional distributions are standard probability distributions \citep{MS15}. By introducing latent variables $\psi$ and $\nu_i$ for $i=k+2,\dots,n$, it holds that
\begin{align*}
& \tau\sim C^+(0,1) \iff \tau^2\mid \psi \sim \mathrm{IG}(1/2, 1/\psi), \ \ \psi\sim \mathrm{IG}(1/2,1), \\
&u_i\sim C^+(0,1) \iff u_i^2\mid \nu_i \sim \mathrm{IG}(1/2, 1/\nu_i), \ \ \nu_i\sim \mathrm{IG}(1/2,1),\quad i=k+2,\dots,n.
\end{align*}
Then we have the following Markov chain Monte Carlo algorithm under horseshoe prior. We note that the full conditional distributions of $\theta$ and $\sigma^2$ depend on whether we assume the shape constraint or not. Furthermore, when we consider the shape constraint, we need to sample additional parameters $\rho^2$ and $v_i$ for $i=1,\dots, n-1$ from the posterior. 

\subsubsection*{Gibbs sampling algorithm under horseshoe prior}

\begin{enumerate}
\item Sampling of $\theta$
\begin{itemize}
\item Draw $\omega_i \sim \mathrm{PG}(1, \eta e_i^{\top} \xi)=\mathrm{PG}(1,\eta\xi_i)$,  independently for $i=1,\dots,n$.
\item Draw $\xi\sim N_n (\mu_{\omega}, \Sigma_{\omega})$, with
\begin{align*}
&\Sigma_{\omega}=\left(\eta^2 \Omega + A\right)^{-1},\quad \mu_{\omega}=\Sigma_{\omega}\left(\eta \kappa +b\right),\\
&\kappa=(1/2,\dots,1/2)^{\top}, \quad \Omega=\mathrm{diag}(\omega_1,\dots,\omega_n),
\end{align*}
where $A$ and $b$ are as follows.

\begin{itemize}
\item[-] (Unconstraint)
\begin{align*}
A=(I_n+D^{\top}U^{-1}D)/\sigma^2,\quad b=-D^{\top}U^{-1}Dy.
\end{align*}
\item[-] (Nearly isotonic constraint)
\begin{align*}
A=(I_n+D^{\top}U^{-1}D+PVP)/\sigma^2,\quad b=-(D^{\top}U^{-1}Dy+P^{\top}V(Py+v)),
\end{align*}
where $P=D^{(1)}$ and $V=\mathrm{diag}(1/(2\rho^2 v_1),\dots, 1/(2\rho^2v_{n-1}))$.
\end{itemize}
\item Set $\theta=\xi+y$
\end{itemize}
\item Sampling of $\sigma^2$
\begin{itemize}
\item Draw $\sigma^2 \sim\mathrm{IG}\left(\alpha_{\sigma}, \beta_{\sigma}\right)$, where $\alpha_{\sigma}$ and $\beta_{\sigma}$ are as follows.
\begin{itemize}
\item[-] (Unconstraint) 
\begin{align*}
\alpha_{\sigma}=n+a_{\sigma},\quad \beta_{\sigma}=\sum_{i=1}^n(y_i-\theta_i)^2/2+\theta^{\top}D^{\top}U^{-1}D\theta/2+b_{\sigma}.
\end{align*}
\item[-] (Nearly isotonic constraint)
\begin{align*}
\alpha_{\sigma}&=(3n-1)/2+a_{\sigma},\\
 \beta_{\sigma}&=\sum_{i=1}^n(y_i-\theta_i)^2/2+\theta^{\top}D^{\top}U^{-1}D\theta/2+\sum_{i=1}^{n-1} (P\theta)_i^2/(2\rho^2v_i)+b_{\sigma}.
\end{align*}
\end{itemize}
\end{itemize}
\item Sampling of $\tau$
\begin{itemize}
\item Draw $\psi \sim \mathrm{IG}\left(1/2,1/\tau^2+1\right)$.
\item Draw $\tau^2 \sim \mathrm{IG}\left((n-k)/2, \sum_{i=k+2}^{n}(D\theta)_i^2/(2\sigma^2)+1/\psi\right)$.
\end{itemize}
\item Sampling of $u$
\begin{itemize}
\item Draw $u_i^2\sim \mathrm{IG}\left(1+a_{u_i},(D\theta)_i^2/(2\sigma^2)+ b_{\sigma}\right)$, independently for $i=1,\dots,k+1$.
\item Draw $\nu_i^2\sim \mathrm{IG}\left(1/2,1/u_i+1\right)$, independently for $i=k+2,\dots,n$.
\item Draw $u_i^2 \sim\mathrm{IG}\left(1, (D\theta)_i^2/(2\sigma^2\tau^2)+1/\nu_i\right)$, independently for $i=k+2,\dots,n$.
\end{itemize}
\item Sampling of $\rho^2$ {\bf (only when we assume the nearly isotonic constraint)}
\begin{itemize}
\item Draw $v_i \sim \mathrm{GIG}\left(1/2, (P\theta)_i^2/(4\rho^2\sigma^2),1/2\rho^2\sigma^2 \right)$, independently for $i=1,\dots,n-1$.
\item Draw $\rho^2 \sim \mathrm{IG}\left((n-1)/2+a_{\rho}, \sum_{i=1}^{n-1} (P\theta)_i^2/(2\sigma^2v_i)+b_{\rho}\right)$.
\end{itemize}
\end{enumerate}

In practical use, we specify hyper-parameters $a_{\sigma}$, $b_{\sigma}$, $a_{\rho}$, $b_{\rho}$, $a_{u_i}$, $b_{u_i}$ for $i=1,\dots, k+1$. In our numerical studies, we set hyper-parameters $a_{\sigma}= b_{\sigma}=0.1$, $a_{\rho}=b_{\rho}=1$, and $a_{u_i}=b_{u_i}=1$ for $i=1,\dots, k+1$ as default values.

\section{Numerical experiments}
\label{sec:3}

We illustrate the performance of the proposed method through simulation studies. In Subsection \ref{subsec:3.1}, we deal with monotone functions as true regression functions. The aim is to compare the proposed methods with frequentist methods and to see the difference between unconstrained  methods and nearly isotonic constraint methods. In Subsection \ref{subsec:3.2}, we will discuss the robustness of the shape constraint methods (nearly isotonic or monotone constraint) for structural misspecification such that the true boundary trend is not completely monotone.

\subsection{Simulation (I): Estimation of monotone boundary}
\label{subsec:3.1}

We generate the data from the model $y_i=f(x_i)+\varepsilon_i$ ($i=1,\dots,100$), where $f(x)$ and $\varepsilon$ are a true function and a noise distribution, respectively. We assume the following two true functions:
\begin{itemize}
\item[(i)] Square root (Sqrt): $f(x)=\sqrt{x}/2$ 
\item[(ii)] Piecewise constant (PC):
\begin{align*}
        f(x)=0.5\cdot 1_{[1,20]}(x)+1_{[21,40]}(x)
        +2.5\cdot I_{[41,60]}(x)+3.5\cdot 1_{[61,100]}(x)
        \end{align*}
\end{itemize}
We employ the (upper truncated) half-normal distribution $\mathrm{HN}(0,\sqrt{\pi/2}/\sigma)$ with location parameter $0$ and variance parameter $\sigma^2$ to generate the noise $\varepsilon_i$. We consider the four scenarios: (a) $\sigma=0.5$, (b) $\sigma=1$, (c) $\sigma=2$ and (d) mixtures of the half-normal $0.8\times\mathrm{HN}(0,\sqrt{\pi/2})+0.2\times\mathrm{HN}(0,\sqrt{\pi/2}/3)$. Hereafter, we often denote scenarios like (i-a) for example. We adopt the proposed methods: Bayesian boundary trend filtering under the horseshoe, Laplace, and normal priors (denoted by HS, Lap, and Nor) and nearly isotonic (NI) constrained Bayesian boundary trend filtering for each prior (denoted by HSNI, LapNI, and NorNI for short). As competitors, we consider the following frequentist methods:
\begin{itemize}
\item QS, CS, QSI, and CSI: Quadratic and Cubic spline methods (with/without isotonic constraint) for estimating boundary curve proposed by \cite{DNP16}. The knot of the spline is selected by Bayesian information criterion (BIC). R-code is provided by \texttt{R} package \texttt{npbr}. 
\item QTF: Quantile trend filtering method proposed by \cite{BGC20}. The method solves the optimization problem using the alternating direction method of multipliers (ADMM) algorithm, where the penalty parameter is determined by the extended Bayesian information criterion (eBIC). We use the quantile level $0.99$ to estimate the extremal quantile trend. The method can be implemented by using their {\tt R} package\footnote{\url{https://github.com/halleybrantley/detrendr}}.
\item FDH and LFDH: Classical nonparametric methods called the free disposal hull defined by \cite{DST84} and its linearized version \citep[see e.g.][]{HP02}. The method can also be implemented by \texttt{R} package \texttt{npbr}.
\end{itemize}

For the proposed Bayesian methods, we generated 10500 posterior samples, then the we removed first 500 samples, and only every 5th scan was saved. For trend filtering methods (including QTF), we set the order of $k$ as $k=1$ (piecewise linear) and $k=0$ (piecewise constant) for scenarios (i) and (ii) respectively. For the proposed method, we set $\eta=500$.

To evaluate the performance of estimates, we adopted the root mean squared error (RMSE), the average length of the credible interval (AL), and the coverage probability (CP). These criteria  are defined by $\mathrm{RMSE}=\{n^{-1}\sum_{i=1}^n (f(x_i)-\hat{\theta}_i)^2\}^{1/2}$, $\mathrm{AL}=n^{-1}\sum_{i=1}^n (\hat{\theta}_{97.5,i}-\hat{\theta}_{2.5,i})$, and $\mathrm{CP}=n^{-1}\sum_{i=1}^n 1_{[\hat{\theta}_{2.5,i},\hat{\theta}_{97.5,i}]}(f(x_i))$, respectively, 
where $\hat{\theta}_{100(1-\alpha),i}$ represents the $100(1-\alpha)$\% posterior quantile of $\theta_i$. These values were averaged over 100 replications of simulating datasets. We only reported RMSE for frequentist competitors. 

\begin{figure}[htpb]
\centering
\includegraphics[width=15cm]{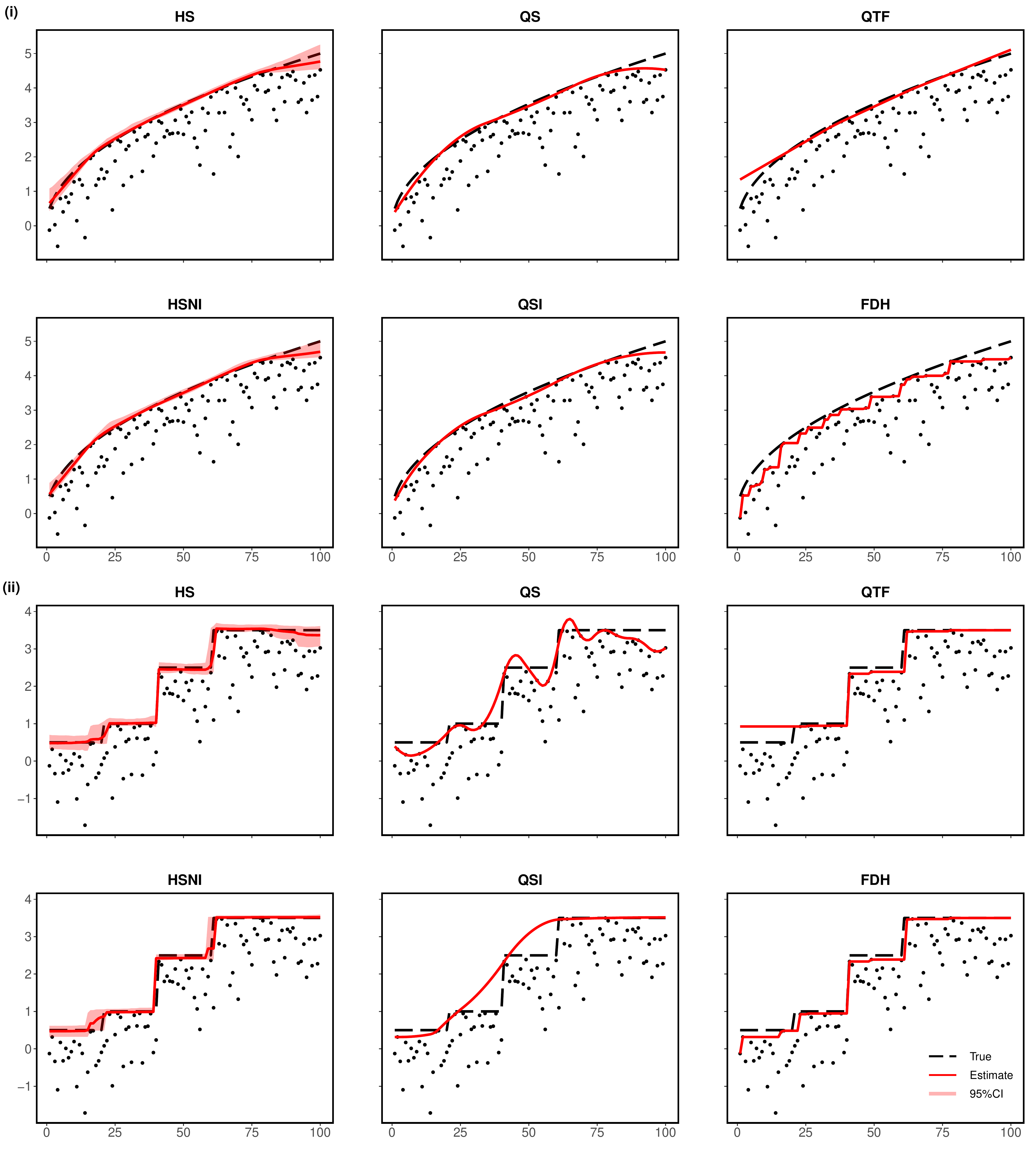}
\caption{One-shot simulation results of six methods for two scenarios under the noise (b). For (i) Sqrt scenario (top panels) and (ii) PC scenario (bottom panels), the resulting estimates of HS, HSNI, QS, QSI, QTF, and FDH are shown.}
\label{sim-oneshot}
\end{figure}

First, we show one-shot simulation results for some methods in Figure \ref{sim-oneshot} when (b) $\sigma=1$. We can observe that five methods (HS, HSNI, QS, QSI, and QTF) give reasonable estimates under scenario (i). For scenario (ii), the QS and QSI methods provide over-shrinkage estimates and the FDH and QTF methods can not capture some change points, while the proposed methods under the horseshoe prior give reasonable estimates for piecewise constant structure. A remarkable point is that the proposed methods under the horseshoe prior illustrate good performance for both scenarios. 

We also report RMSE, AL, and CP averaged over 100 Monte Carlo replications in Tables \ref{table-RMSE} and \ref{table-ALCP}. 
From Table \ref{table-RMSE}, the results indicate that the proposed HS and HSNI methods provide more accurate point estimates than other Bayesian and frequentist methods except for scenarios (i-c) and (i-d). Focusing on the scenario (i-d), we can observe that spline methods provide relatively smaller RMSE than those of the proposed Bayesian methods, and the proposed shape constrained methods significantly improve the RMSE of unconstraint methods. From Table \ref{table-ALCP}, the coverage probabilities under the HS and HSNI methods are larger than the nominal level of 0.95 except for a few cases, whereas the average length of intervals of the HS and HSNI methods tend to be smaller than other methods under scenario (ii). Although the proposed methods under the Laplace prior also have reasonable coverage probabilities, the average lengths of interval tend to be wider than those of the horseshoe prior. 

\begin{table}[htbp]
\caption{RMSE averaged over 100 Monte Carlo replications and their standard deviation (shown in parenthesis). The best score is bolded.}
\begin{center}
\begin{tabular}{c|cccc}
 \toprule
   \multicolumn{5}{c}{(i) Sqrt}\\
  \hline
   & (a) & (b) & (c) & (d) \\ 
   \hline
HS & {\bf0.041} (0.013) & 0.072 (0.027) & 0.122 (0.050) & 0.257 (0.141) \\ 
  Lap & 0.134 (0.026) & 0.265 (0.058) & 0.528 (0.100) & 0.347 (0.088) \\ 
  Nor & 0.049 (0.011) & 0.078 (0.025) &  {\bf0.121} (0.050) & 0.149 (0.031) \\ 
  HSNI &  {\bf0.041} (0.014) &  {\bf0.070} (0.028) & 0.131 (0.075) & 0.174 (0.047) \\ 
  LapNI & 0.090 (0.023) & 0.153 (0.039) & 0.308 (0.083) & 0.154 (0.043) \\ 
  NorNI & 0.047 (0.012) & 0.077 (0.026) & 0.122 (0.056) & 0.138 (0.027) \\ 
  QS & 0.083 (0.043) & 0.133 (0.088) & 0.231 (0.173) & 0.120 (0.055) \\ 
  CS & 0.078 (0.043) & 0.130 (0.082) & 0.248 (0.179) & 0.125 (0.062) \\ 
  QSI & 0.064 (0.026) & 0.094 (0.044) & 0.166 (0.091) &  {\bf0.108} (0.052) \\ 
  CSI & 0.060 (0.027) & 0.092 (0.042) & 0.171 (0.087) & 0.109 (0.050) \\ 
  QTF & 0.145 (0.037) & 0.136 (0.041) & 0.170 (0.053) & 0.445 (0.042) \\ 
  FDH & 0.209 (0.025) & 0.310 (0.042) & 0.488 (0.100) & 0.357 (0.080) \\ 
  LFDH & 0.170 (0.033) & 0.363 (0.120) & 0.876 (0.322) & 0.952 (0.402) \\ 
  \midrule
  \midrule
     \multicolumn{5}{c}{(ii) PC}\\
  \hline
 & (a) & (b) & (c) & (d) \\ 
  \hline
HS &  {\bf0.060} (0.026) & 0.131 (0.041) & 0.237 (0.058) & 0.276 (0.163) \\ 
  Lap & 0.168 (0.025) & 0.297 (0.051) & 0.583 (0.116) & 0.395 (0.112) \\ 
  Nor & 0.163 (0.016) & 0.221 (0.027) & 0.282 (0.050) & 0.265 (0.040) \\ 
  HSNI & 0.061 (0.035) &  {\bf0.120} (0.064) &  {\bf0.230} (0.077) &  {\bf0.174} (0.070) \\ 
  LapNI & 0.122 (0.017) & 0.193 (0.030) & 0.318 (0.069) & 0.235 (0.038) \\ 
  NorNI & 0.273 (0.023) & 0.292 (0.032) & 0.291 (0.033) & 0.310 (0.038) \\ 
  QS & 0.227 (0.034) & 0.320 (0.059) & 0.459 (0.147) & 0.349 (0.084) \\ 
  CS & 0.240 (0.034) & 0.334 (0.059) & 0.468 (0.159) & 0.364 (0.089) \\ 
  QSI & 0.260 (0.079) & 0.333 (0.076) & 0.377 (0.059) & 0.352 (0.068) \\ 
  CSI & 0.286 (0.083) & 0.364 (0.054) & 0.365 (0.063) & 0.364 (0.061) \\ 
  QTF & 0.207 (0.029) & 0.251 (0.050) & 0.370 (0.223) & 0.273 (0.057) \\ 
  FDH & 0.141 (0.032) & 0.239 (0.068) & 0.414 (0.108) & 0.302 (0.089) \\ 
  LFDH & 0.350 (0.079) & 0.602 (0.163) & 1.297 (0.403) & 1.293 (0.412) \\
  \bottomrule
\end{tabular}
\label{table-RMSE}
\end{center}
\end{table}

\begin{table}[htbp]
\caption{Average lengths (AL) and coverage probabilities (CP) of 95\% credible intervals averaged over 100 Monte Carlo replications.}
\begin{center}
\begin{tabular}{c|cc|cc|cc|cc}
  \toprule
  \multicolumn{9}{c}{(i) Sqrt}\\
     \hline
 &\multicolumn{2}{c|}{(a)}&\multicolumn{2}{c|}{(b)}&\multicolumn{2}{c|}{(c)}&\multicolumn{2}{c}{(d)}\\
   \hline
 & AL & CP & AL & CP & AL & CP & AL & CP \\ 
  \hline
HS & 0.200 & 0.990 & 0.342 & 0.983 & 0.731 & 0.980 & 0.989 & 0.979 \\ 
  Lap & 0.444 & 0.931 & 0.885 & 0.925 & 1.747 & 0.925 & 1.588 & 0.975 \\ 
  Nor & 0.175 & 0.969 & 0.271 & 0.961 & 0.620 & 0.972 & 0.853 & 0.990 \\ 
  HSNI & 0.183 & 0.987 & 0.287 & 0.973 & 0.413 & 0.939 & 0.637 & 0.955 \\ 
  LapNI & 0.261 & 0.919 & 0.419 & 0.890 & 0.778 & 0.860 & 0.638 & 0.972 \\ 
  NorNI & 0.160 & 0.960 & 0.219 & 0.920 & 0.343 & 0.901 & 0.430 & 0.956 \\
   \midrule
      \midrule
  \multicolumn{9}{c}{(ii) PC}\\
     \hline
 &\multicolumn{2}{c|}{(a)}&\multicolumn{2}{c|}{(b)}&\multicolumn{2}{c|}{(c)}&\multicolumn{2}{c}{(d)}\\
   \hline
 & AL & CP & AL & CP & AL & CP & AL & CP \\ 
  \hline
HS & 0.211 & 0.980 & 0.438 & 0.969 & 0.948 & 0.958 & 0.780 & 0.975 \\ 
  Lap & 0.564 & 0.942 & 1.024 & 0.941 & 1.927 & 0.935 & 1.616 & 0.961 \\ 
  Nor & 0.583 & 0.944 & 0.813 & 0.939 & 1.259 & 0.957 & 0.978 & 0.929 \\ 
  HSNI & 0.121 & 0.936 & 0.210 & 0.911 & 0.380 & 0.834 & 0.340 & 0.938 \\ 
  LapNI & 0.301 & 0.872 & 0.472 & 0.854 & 0.814 & 0.844 & 0.682 & 0.901 \\ 
  NorNI & 0.329 & 0.698 & 0.404 & 0.700 & 0.533 & 0.744 & 0.494 & 0.714 \\ 
   \bottomrule
\end{tabular}
\label{table-ALCP}
\end{center}
\end{table}

\subsection{Simulation (II): Estimation of piecewise monotone boundary}
\label{subsec:3.2}
In this subsection, we give an additional simulation study to show the usefulness of the nearly isotonic constraint. Although the monotone constraint $\theta_1\le \cdots \le \theta_n$ is not robust against structural misspecification, we show that the proposed nearly isotonic method provides a reasonable estimate even if the monotonicity is partially violated. To verify the robustness of the proposed method, we consider a piecewise monotone function as a true function in the data generating process $y_i=f(x_i)+\varepsilon_i$ for $x_i=1,\dots,100$. To this end, we consider the following piecewise sigmoid function \citep[see also][]{MW00, M20}:
\begin{align*}
        f(x)&=1+4\exp(32x-8)/(1+\exp(16x-8))\cdot 1_{[0,50]}(x)\\
        & \ \ \ \ 
        +4\exp(16(2x-1)-8)/(1+\exp(16(2x-1)-8))\cdot 1_{[51,100]}(x).
\end{align*}
The true function is monotone except for a jumping point at $x=50$, then the scenario is reasonable to compare the HS, HSNI, and other isotonic constraint methods. As noise distributions, we also assume the half-normal distribution $\mathrm{HN}(0,\sqrt{\pi/2}/\sigma)$ with (a) $\sigma=0.5$, (b) $\sigma=1$ and (c) $\sigma=2$ as in Subsection \ref{subsec:3.1}, 
and report the averaged values of RMSE, AL, and CP. We compare the proposed HS and HSNI methods as well as the frequentist boundary spline methods with/without the isotonic constraint (QS, QSI, CS, and CSI). The number of generated posterior samples is the same as that of Subsection \ref{subsec:3.1}, and we set $k=1$ and $\eta=500$ for the proposed methods. 

\begin{figure}[htpb]
\centering
\includegraphics[width=15cm]{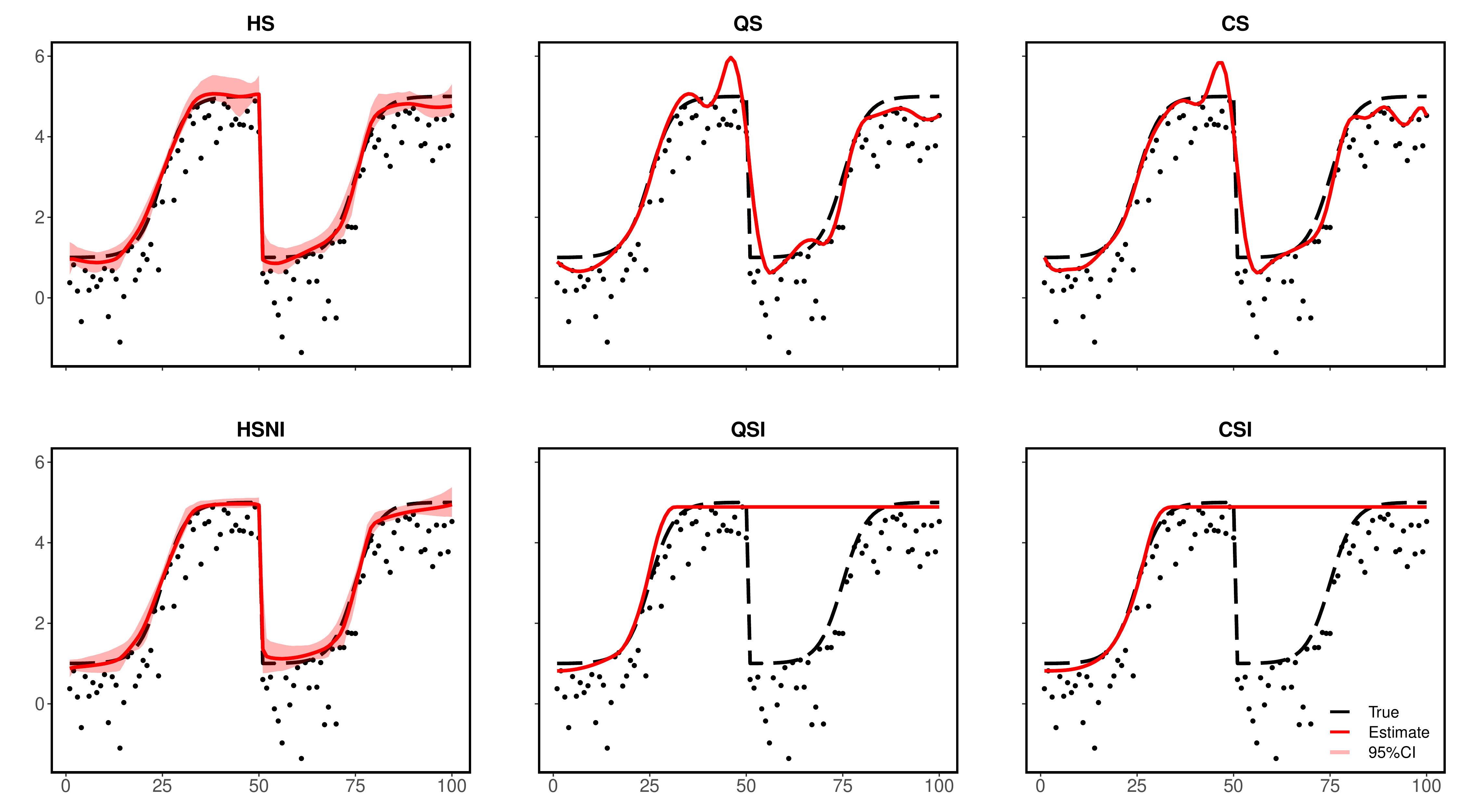}
\caption{One-shot simulation results of six methods (HS, HSNI, QS, QSI, CS, and CSI) for the piecewise sigmoid scenario under the noise (b).}
\label{sim-oneshot-sigmoid}
\end{figure}

We show one-shot simulation results in Figure \ref{sim-oneshot-sigmoid} under case (b). From the figure, it is observed that the QS and CS methods can not capture well the jump point at $x=50$. The QSI and CSI do not work at all because of model misspecification. In contrast to such methods, the proposed HS and HSNI methods provide smoother trends and their estimates seem to be comparable. The result shows that the proposed HSNI method is more robust against structural misspecification than that of assuming a completely monotone constraint. 

From Table \ref{table-sim2-RMSE}, it is observed that the HSNI method has the smallest RMSE. For uncertainty quantification, we report average lengths and coverage probabilities of credible intervals in Table \ref{table-sim2-ALCP}. The proposed methods have reasonable coverage probabilities except for the HSNI under (c). Therefore, the proposed nearly isotonic method can reasonably estimate a piecewise monotone trend as well as a completely monotonic trend.

\begin{table}[htbp]
\caption{RMSE averaged over 100 Monte Carlo replications and their standard deviation (shown in parenthesis). The best score is bolded.}
\begin{center}
\begin{tabular}{c|ccc}
  \toprule
       & (a) & (b) & (c) \\ 
  \hline
HS & 0.089 (0.026) & 0.194 (0.065) & 0.408 (0.083) \\ 
  HSNI & {\bf0.082} (0.019) & {\bf0.178} (0.064) & {\bf0.375} (0.094) \\ 
  QS & 0.467 (0.074) & 0.515 (0.093) & 0.659 (0.110) \\ 
  CS & 0.478 (0.065) & 0.536 (0.076) & 0.706 (0.103) \\ 
  QSI & 1.826 (0.017) & 1.809 (0.030) & 1.790 (0.052) \\ 
  CSI & 1.826 (0.019) & 1.808 (0.031) & 1.794 (0.055) \\ 
    \bottomrule
\end{tabular}
\label{table-sim2-RMSE}
\end{center}
\end{table}

\begin{table}[htbp]
\caption{Average lengths (AL) and coverage probabilities (CP) of 95\% credible intervals averaged over 100 Monte Carlo replications.}
\begin{center}
\begin{tabular}{c|cc|cc|cc}
  \toprule
 &\multicolumn{2}{c|}{(a)}&\multicolumn{2}{c|}{(b)}&\multicolumn{2}{c}{(c)}\\
   \hline
 & AL & CP & AL & CP & AL & CP \\ 
  \hline
HS & 0.363 & 0.967 & 0.645 & 0.961 & 1.205 & 0.933 \\ 
  HSNI & 0.311 & 0.959 & 0.495 & 0.927 & 0.812 & 0.868 \\ 
   \bottomrule
\end{tabular}
\label{table-sim2-ALCP}
\end{center}
\end{table}

\subsection{Sensitivity analysis for selecting of $\eta$}
\label{subsec:3.3}

\begin{figure}[htpb]
\centering
\includegraphics[width=12cm]{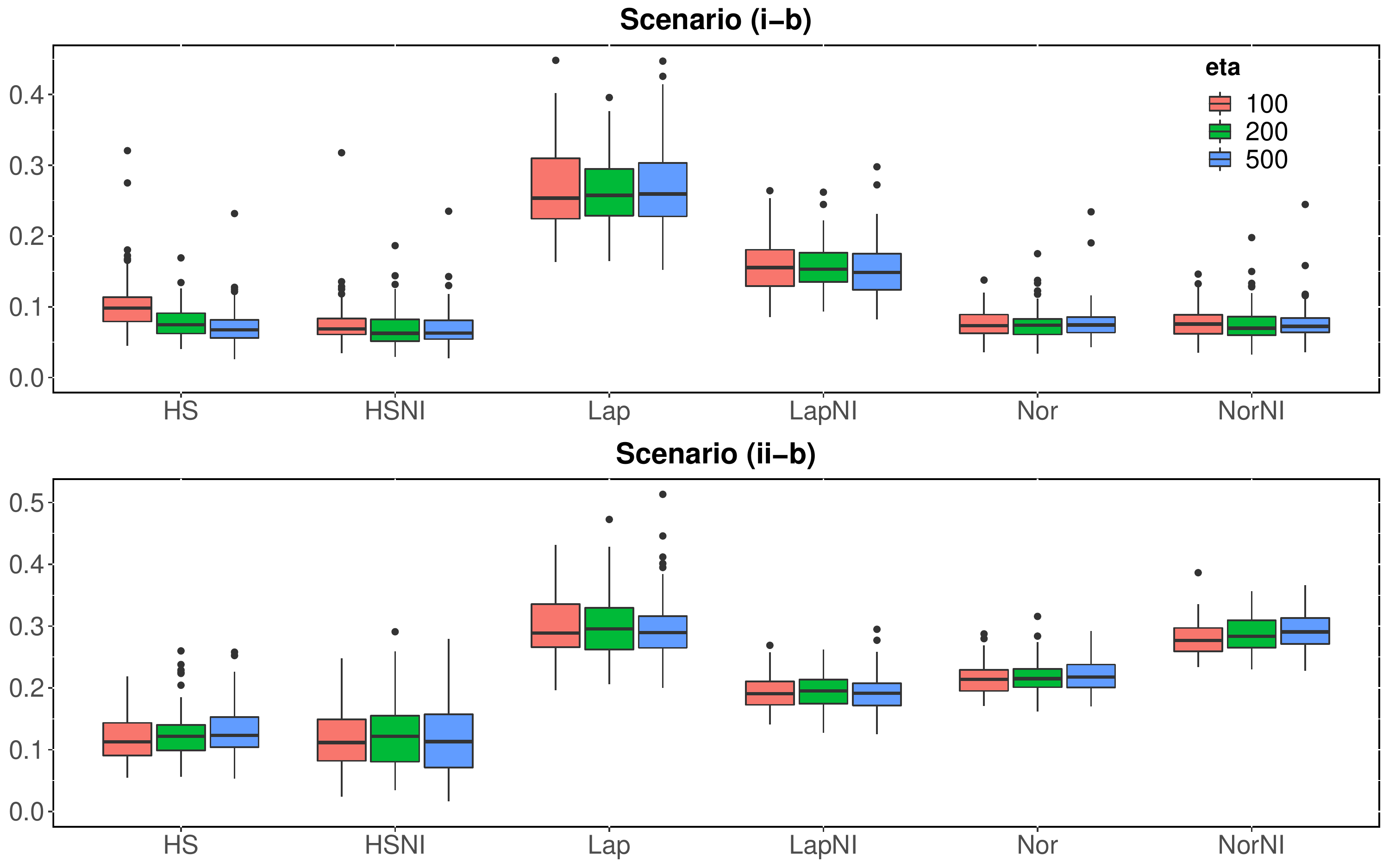}
\caption{Sensitivity analysis of $\eta$ under scenarios (i-b) and (ii-b).}
\label{boxplot_RMSE_eta}
\end{figure}

Since the sigmoid function $\sigma_{\eta}(x)$ defined by \eqref{sigmoid} converges in the sense of $L_1$-convergence to the indicator function $1_{[0,\infty)}(x)$ as $\eta\to\infty$, we may select a moderate large $\eta$ in practice. We here check the sensitivity of the point estimates of $\theta$ for various values of $\eta$ using the same simulated dataset as Subsection \ref{subsec:3.1}. We considered three values $\eta \in \{100,200,500\}$. The boxplots of RMSE for the two scenarios (i) Sqrt and (ii) PC are provided in Figure \ref{boxplot_RMSE_eta}, where the noise distribution is the half-normal with the standard deviation $\sigma=1$. From these figures, we can observe that the results of RMSE for each $\eta$ do not change very much.

\subsection{Efficiency of sampling}
\label{subsec:3.4}

We evaluate the efficiency of the proposed Gibbs sampler under the same simulation setting as in Subsection \ref{subsec:3.1}. We here adopt scenario (i-a) in Subsection \ref{subsec:3.1} and employ the HS method. 

The rejection sampler proposed by \cite{B17} is known as an efficient sampling method from the tMVN distribution, and it is interesting to compare the proposed sampler and Botev's one. However, the algorithm gets stack even if the dimension $n$ is equal to $50$ in our model because the acceptance rate becomes very low in high-dimension \citep[see also][]{SBP18}. \cite{pakman2014} also proposed an excellent sampling algorithm based on Hamiltonian Monte Carlo, and the algorithm can be easily implemented by using their R-package {\tt tmg}. However, it is known that their method has the following problems: 1) Leaf-frog steps with careful tuning are necessary to obtain good mixing; 2) The algorithm often fails to produce answers and has high-computational cost \citep[see also][]{SBP18, RPB20}. For these reasons, we considered a coordinate-wise sampler to obtain samples from the tMVN distribution as a competitor. Since the full conditional distribution of parameter vector $\theta=(\theta_1,\dots,\theta_n)$ is the tMVN distribution, we can easily derive the full conditional distribution of $\theta_i$ given $\theta_{-i}=(\theta_1,\dots,\theta_{i-1},\theta_{i+1},\dots,\theta_n)$ as one-dimensional truncated normal distribution \citep[e.g.][]{OHIS22}. By using the coordinate-wise sampler, we can construct the Gibbs sampler in our model without approximating the indicator function. Although the run-time of the coordinate-wise sampler is faster than that of the proposed Gibbs sampler, especially for high-dimensional situation, we show that the proposed method is more efficient in terms of the effective sample size (ESS) through a simple simulation study. 

\begin{figure}[htpb]
\centering
\includegraphics[width=12cm]{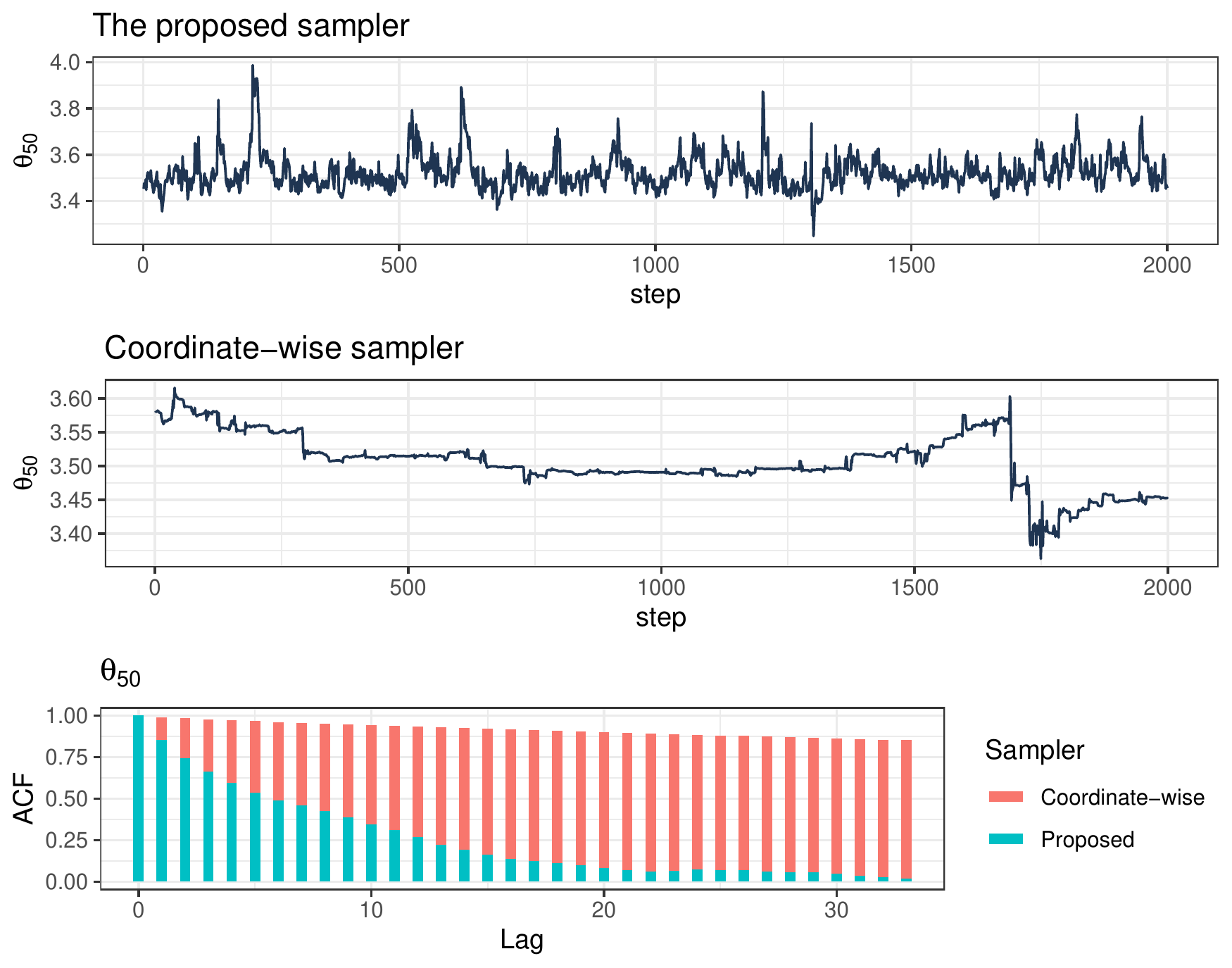}
\caption{Trace plots and autocorrelations of posterior samples of $\theta_{50}$ under scenario (i-b) (in Subsection \ref{subsec:3.1}) with HS method. We used 2000 posterior samples after a burn-in period of 500 and after thinning the chain by 5.}
\label{fig:mcmc-plot}
\end{figure}

We report the result of the sampling efficiency of two methods in Table \ref{table:efficiency}. For a low dimensional case such as $n=50$, they are comparable. On the other hand, it is observed that the larger dimension, the larger difference between them. For the coordinate-wise sampler, although run-time is relatively faster than the proposed method, the ESS is relatively small against the proposed one, especially for high-dimension. The run-time of the proposed method increases rapidly with dimension, but the ESS is relatively better than that of the coordinate-wise sampler. Interestingly, we can observe that the ESS of the proposed method tends to increase with dimension. As an example, we show the sample path and autocorrelation plot of parameter $\theta_{50}$ which is a specific location in Figure \ref{fig:mcmc-plot}. It indicated that the autocorrelation does not rapidly decay for the coordinate-wise sampler, while the proposed method has reasonable mixing and autocorrelation.

\begin{table}[htbp]
\caption{CPU times (in seconds) and the mean of effective sample sizes (ESS) for $\theta_1,\dots,\theta_n$ under scenario (i-b) (in Subsection \ref{subsec:3.1}) with the HS method, averaged over 10 runs which are generated 2000 posterior samples after a burn-in period of 500 and after thinning the chain by 5.}
\begin{center}
\begin{tabular}{c|cc|cc|cc|cc}
  \toprule
 &\multicolumn{2}{c|}{$n=50$}&\multicolumn{2}{c|}{$n=100$}&\multicolumn{2}{c|}{$n=200$}&\multicolumn{2}{c}{$n=300$}\\
   \hline
 & Time & ESS & Time & ESS & Time & ESS & Time & ESS \\ 
  \hline
Proposed & 11.17 & 69.75 & 47.73 & 92.59 & 289.14 & 112.19 & 910.72 & 112.07  \\ 
  Coordinate-wise & 12.12 & 11.83 & 32.99 & 11.98 & 128.34 & 9.35 & 343.50 & 9.57 \\ 
   \bottomrule
\end{tabular}
\label{table:efficiency}
\end{center}
\end{table}

\section{Real data examples}
\label{sec:4}

We apply the proposed methods to two real data examples. 

\subsection{Production activity of air traffic controllers}
\label{subsec:4.1}

\begin{figure}[htpb]
\centering
\includegraphics[width=12cm]{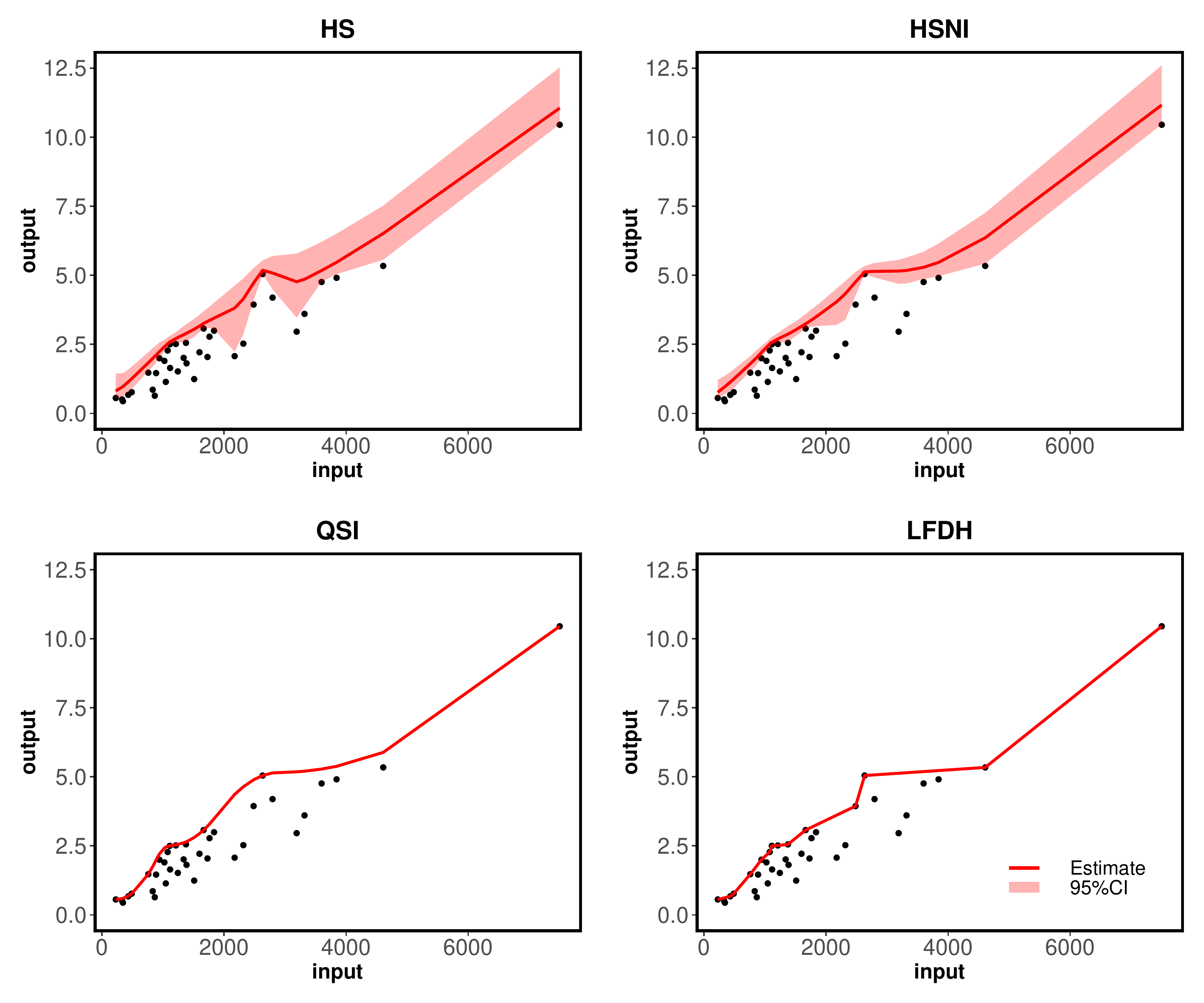}
\caption{Estimated boundary curves for the proposed methods (HS and HSNI) and existing 
frequentist methods (QSI and LFDH) for air controllers data.}
\label{air_plot}
\end{figure}

We consider an efficient frontier estimation that corresponds to the production activity of the 37 European air traffic controllers \citep{MS02}. The data was also analyzed by \cite{DNP16}, and they applied their boundary spline methods. We can obtain the data in {\tt R} package {\tt npbr}. From the scatter plot of the data in Figure \ref{air_plot}, we observe that the assumption of a monotone boundary seems to be reasonable. The $x$ and $y$-axis indicate the input (an aggregate factor of a different kind of labor) and output (an aggregate factor of the activity produced, based on the number of controlled air movements, the number of controlled flight hours, etc.) variables. We applied the proposed HS and HSNI methods compared with existing boundary quadratic spline with isotonic constraint (QSI) and LFDH methods. We note that Brantley's quantile trend filtering method we used in the previous section can not be applied to irregular grid data. To handle such an irregular grid, we need to use the proposed trend filtering methods with an adjusted difference matrix defined in Remark \ref{rem2.1}. To avoid numerical instability of the matrix $D^{\top}U^{-1}D$ in the proposed methods, we employed a transformation of the input variables as $1000\times x$. We set the order of trend filtering as $k=1$ and generated 10500 samples (burn-in 500), and then saved the 5th scanned samples.

The results are shown in Figure \ref{air_plot}. Although the proposed HS method provides an almost monotonic point estimate, it is observed a decreasing trend between 2000 and 4000. On the other hand, the proposed HSNI method provides a reasonable point estimate of monotone boundary and uncertainty quantification. The average length of 95\% credible intervals made by the HSNI method was 0.692, which was considerably smaller than the 0.936 produced by the HS method. Although the point estimate using the QSI and HSNI methods seem to be comparable, it is observed that the QSI and LFDH methods tend to give over-fitting estimates to data. In particular, the LFDH method can not estimate a smooth monotone boundary.

\subsection{Global warming data}
\label{subsec:4.2}

\begin{figure}[htpb]
\centering
\includegraphics[width=12cm]{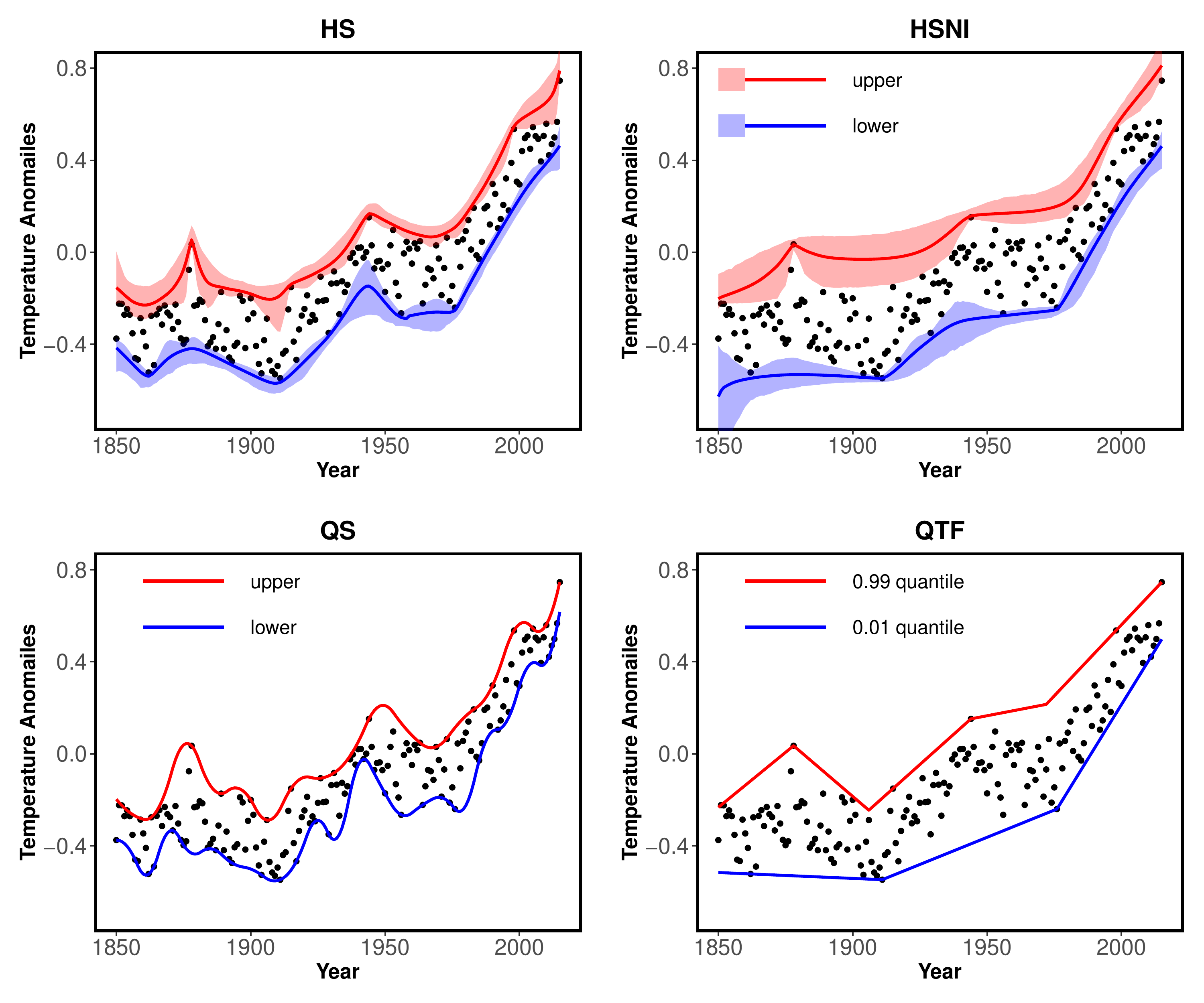}
\caption{Estimated boundary curves for the proposed methods (HS and HSNI) and existing 
frequentist methods (QS and QTF) for global warming data.}
\label{global_temp_upper_lower}
\end{figure}

Global warming is one of the important issues in the world. Although the attention to the problem is often focused on the prediction of future climate change, it is also important to look back and explore the processes of past climate change. We apply the proposed method to estimate the past trend of annual temperature anomalies. The data is available from {\tt R} package {\tt CVXR}, and it includes the global monthly and annual temperature anomalies relative to the mean of 1960--1990 during 1850--2015 (the sample size is 166). The data is equally spaced non-stationary time series data, we can also observe that the monotonicity assumption is partially violated from the scatter plot in Figure \ref{global_temp_upper_lower}. There exist a few outliers in 1877 and 1878, and they are reported as unexpected climate change \citep[see e.g.][]{ClimateChange}. Hence, the use of the nearly isotonic constraint may be useful. For the data, \cite{THT11} applied the nearly isotonic regression to estimate the mean trend. In our analysis, we are interested in estimation of the (upper and lower) boundary trends not in the mean trend. Estimating boundary trends is useful to clarify the variability of extreme values, and we can obtain the range of variability as a by-product. 

In this analysis, we applied the proposed HS and HSNI methods. We generated 10500 samples (burn-in 500), and then saved the 5th scanned samples. As competitors, we employ the unconstraint quadratic spline (QS) and quantile trend filtering (QTF). We set quantile levels of the QTF method as 0.99 and 0.01. For the HS, HSNI, and QTF methods, we assume that the order of trend filtering is $k=1$. 

The results of point estimates and credible intervals are shown in Figure \ref{global_temp_upper_lower}. It is observed that the QS method gives a slightly overfitted estimate of the data and quantile trend filtering tends to induce over-shrinkage. On the other hand, the proposed HS and HSNI methods provide smoother and more locally adaptive boundary trend estimates. The HSNI method provides almost monotone boundary trend estimates, while for the upper boundary, the monotonicity violates during 1878--1920. Furthermore, average distances between upper and lower point estimates which is defined by $\sum_{i=1}^{166} |\hat{\theta}_{\mathrm{upper},i} - \hat{\theta}_{\mathrm{lower},i}|/166$ were 0.328 (HS), 0.439 (HSNI), 0.269 (QS), and 0.431 (QTF), respectively. In terms of the average distance, the proposed HSNI method is quite similar to the QTF method, and it is also considered that the QS method overfits the data and underestimates the extent to which data exists. 
For uncertainty quantification, average lengths of 95\% credible intervals made by the HS method were 0.100 (lower) and 0.117 (upper), which are smaller than 0.126 (lower) and 0.164 (upper) by the HSNI method, respectively. At first glance, this result appears as if the assumption of shape constraint was not a reasonable one, but it also appears to successfully capture the uncertainty in the estimation of the shape of the boundary function. For example, in Figure \ref{global_temp_upper_lower}, the 95\% credible intervals for the upper boundary become wider during 1978-1920, and the lower and upper limits of the intervals give the trends like the HS method and monotone trend, respectively. Hence, it indicates that the HSNI method provides a reasonable uncertainty evaluation to some extent. As we observed in Section \ref{sec:3}, if the potential boundary is monotone or  not monotone, the resulting 95\% credible intervals of the HSNI method tend to be narrower than those of the HS method, unlike this result.

\section{Concluding remarks}

In this paper, we proposed a Bayesian boundary trend filtering using the truncated multivariate normal working likelihood and global-local shrinkage priors. Using the approximation of indicator function in the truncated multivariate normal likelihood, an efficient Gibbs sampling algorithm to sample from posterior distribution was also constructed. 

We close this paper by considering some future directions. Although we employ the truncated multivariate normal distribution as a working likelihood, it may lead to undesirable inference when the model is misspecified. To overcome this problem, we need to consider a kind of calibration method to obtain the correct coverage probability of credible interval \citep[e.g.][]{SM19, OHS22fast}. Another important issue is ``robustness" against outliers. However, it is well-known that the robust estimation of boundary curve is not easy unlike mean curve \citep[e.g.][]{DS05, DR06, DFS21}. In our framework, the scale constant $\eta$ in the sigmoid function may play an important role to control the boundary constraint. Developing a suitable selection criterion of $\eta$ in the presence of outliers will be an interesting future work. 

\section*{Acknowledgement}
We thank an Associate Editor and anonymous reviewers for useful suggestions, which improved the quality of this work. This work was supported by JST, the establishment of university fellowships towards the creation of science technology innovation (grant number JPMJFS2129). 
This work is partially supported by the Japan Society for the Promotion of Science (grant number: 21K13835).

\vspace{5mm}
\bibliography{refs}
\bibliographystyle{chicago}

\end{document}